%% file: Paper-33.tex
\documentclass[runningheads]{llncs}
\usepackage[T1]{fontenc}
\usepackage{amssymb}
\usepackage{graphicx,verbatim}
\usepackage{amsmath}
\usepackage{amssymb}
\usepackage{bbding}
\usepackage{pifont}
\usepackage{wasysym}
\usepackage{utfsym}
\usepackage{fontawesome}
\usepackage{url}
\usepackage{mathtools}
\usepackage{color}
\usepackage{xcolor}
\usepackage{soul}
\usepackage[misc]{ifsym}
\usepackage[colorlinks=true,linkcolor=blue,citecolor=blue,urlcolor=blue]{hyperref}

\def\ie{\textit{i.e.}}

\usepackage{multirow}
\begin{document}
\title{Cross-view Generalized Diffusion Model for Sparse-view CT Reconstruction}
%
\author{
Jixiang Chen\inst{1}\orcidID{0000-0001-9941-8324} \and
Yiqun Lin\inst{1}\orcidID{0000-0002-7697-0842} \and 
Yi Qin\inst{1}\orcidID{0009-0000-2236-652X} \and
Hualiang Wang \inst{1}\orcidID{0009-0006-0157-8885} \and
Xiaomeng Li\inst{1(\textrm{\Letter})}\orcidID{0000-0003-1105-8083}
} 
\authorrunning{F. Author et al.}
%
\institute{ Department of Electronic and Computer Engineering, 
The Hong Kong University of Science and Technology, Hong Kong, China \\
\email{eexmli@ust.hk}
}



\maketitle              
\begin{abstract}
Sparse-view computed tomography (CT) reduces radiation exposure by subsampling projection views, but conventional reconstruction methods produce severe streak artifacts with undersampled data. While deep-learning-based methods enable single-step artifact suppression, they often produce over-smoothed results under significant sparsity. Though diffusion models improve reconstruction via iterative refinement and generative priors, they require hundreds of sampling steps and struggle with stability in highly sparse regimes. To tackle these concerns, we present the Cross-view Generalized Diffusion Model (CvG-Diff), which reformulates sparse-view CT reconstruction as a generalized diffusion process. Unlike existing diffusion approaches that rely on stochastic Gaussian degradation, CvG-Diff explicitly models image-domain artifacts caused by angular subsampling as a deterministic degradation operator, leveraging correlations across sparse-view CT at different sample rates. To address the inherent artifact propagation and inefficiency of sequential sampling in generalized diffusion model, we introduce two innovations: Error-Propagating Composite Training (EPCT), which facilitates identifying error-prone regions and suppresses propagated artifacts, and Semantic-Prioritized Dual-Phase Sampling (SPDPS), an adaptive strategy that prioritizes semantic correctness before detail refinement. Together, these innovations enable CvG-Diff to achieve high-quality reconstructions with minimal iterations, achieving \textbf{38.34} dB PSNR and \textbf{0.9518} SSIM for 18-view CT using only \textbf{10} steps on AAPM-LDCT dataset. Extensive experiments demonstrate the superiority of CvG-Diff over state-of-the-art sparse-view CT reconstruction methods. The code is available at {\tt\small \url{https://github.com/xmed-lab/CvG-Diff}}.


\end{abstract}
%
%
%
\input{docs/intro}
\input{docs/method}
\input{docs/experiments}

\input{docs/conclusion}
\bibliographystyle{splncs04}
\bibliography{Paper-33}

\end{document}

%% file: docs/intro.tex
\section{Introduction}
X-ray computed tomography (CT) is crucial in modern clinical diagnosis. However, exposure to ionizing radiation associated with CT scans poses significant health risks~\cite{brenner2007computed}. Consequently, reducing radiation dose while maintaining diagnostic image quality has become a critical challenge in medical imaging research~\cite{miller1983alara}. Sparse-view CT, which minimizes radiation by subsampling projection views during scanning, offers a promising solution. However, conventional reconstruction methods such as filtered back-projection (FBP) produce severe streak artifacts when applied to undersampled data, limiting their clinical utility.
Recent advances in deep learning have shown great potential for addressing sparse-view CT reconstruction~\cite{lin2023learning,lin2024c,lin2024learning,lin2025deepsparse,chen2024spatial}.
Existing methods mainly focus on developing networks in restoring clean CT image from its sparse-view degraded counterpart using paired training data, operating in either the image domain~\cite{ddnet,fbpconv,freeseed,glorei,regformer}, sinogram domain~\cite{lee2017view,lee2018deep,dong2019sinogram}, or both~\cite{dudonet,dudotrans,drone}.
To address the challenge of recovering details from heavily undersampled data, researchers have proposed enhancements such as incorporating advanced network building block~\cite{dudotrans}, adding Fourier domain regularization~\cite{freeseed}, and utilizing feature-level distillation from intermediate-view inputs~\cite{glorei}. Nevertheless, these single-step restoration methods usually produce over-smoothed results.
%
Recent diffusion models leverage generative priors trained on clean CT for iterative posterior sampling, enabling reconstruction across different sparse-view configurations without paired training~\cite{dps,ddb,mcg,sword,cosign,vss,multichannel}, with~\cite{vss} achieving best results comparable to state-of-the-art supervised methods. However, their reliance on hundreds of iterative steps and sensitivity to extreme sparsity regimes remains a critical limitation.
%
%

%
Motivated by these challenges, we propose to leverage the inherent correlation between sparse-view CT scans acquired at varying subsampling rates. Our key insight is that sparse-view images from adjacent subsampling rates exhibit shared semantic features and radial artifact patterns, with higher sampling rates preserving more structural details under milder artifacts. Building on generalized diffusion models~\cite{colddiff}, which enable multi-step restoration by learning inverting arbitrary degradation transformations across severity levels, we introduce the Cross-view Generalized Diffusion Model (CvG-Diff), which reformulates sparse-view CT reconstruction as a generalized diffusion process. Unlike existing diffusion-model-based methods that simulate degradation through Gaussian noise, CvG-Diff explicitly encodes the angular subsampling artifacts as a deterministic degradation operator, enabling unified training across various subsampling rates. In contrast to existing diffusion model~\cite{dps,ddb,mcg,sword,vss,multichannel} and unified model~\cite{yang2025ct,ma2023universal} that focus on accommodating arbitrary subsampling rates to minimize the generalization error, CvG-Diff instead aims to exploit the simultaneous reconstruction capability across various subsampling levels, optimizing the reconstruction quality with minimal iterative steps.

\begin{figure}[t]
    \centering
    \includegraphics[width=1.0\linewidth]{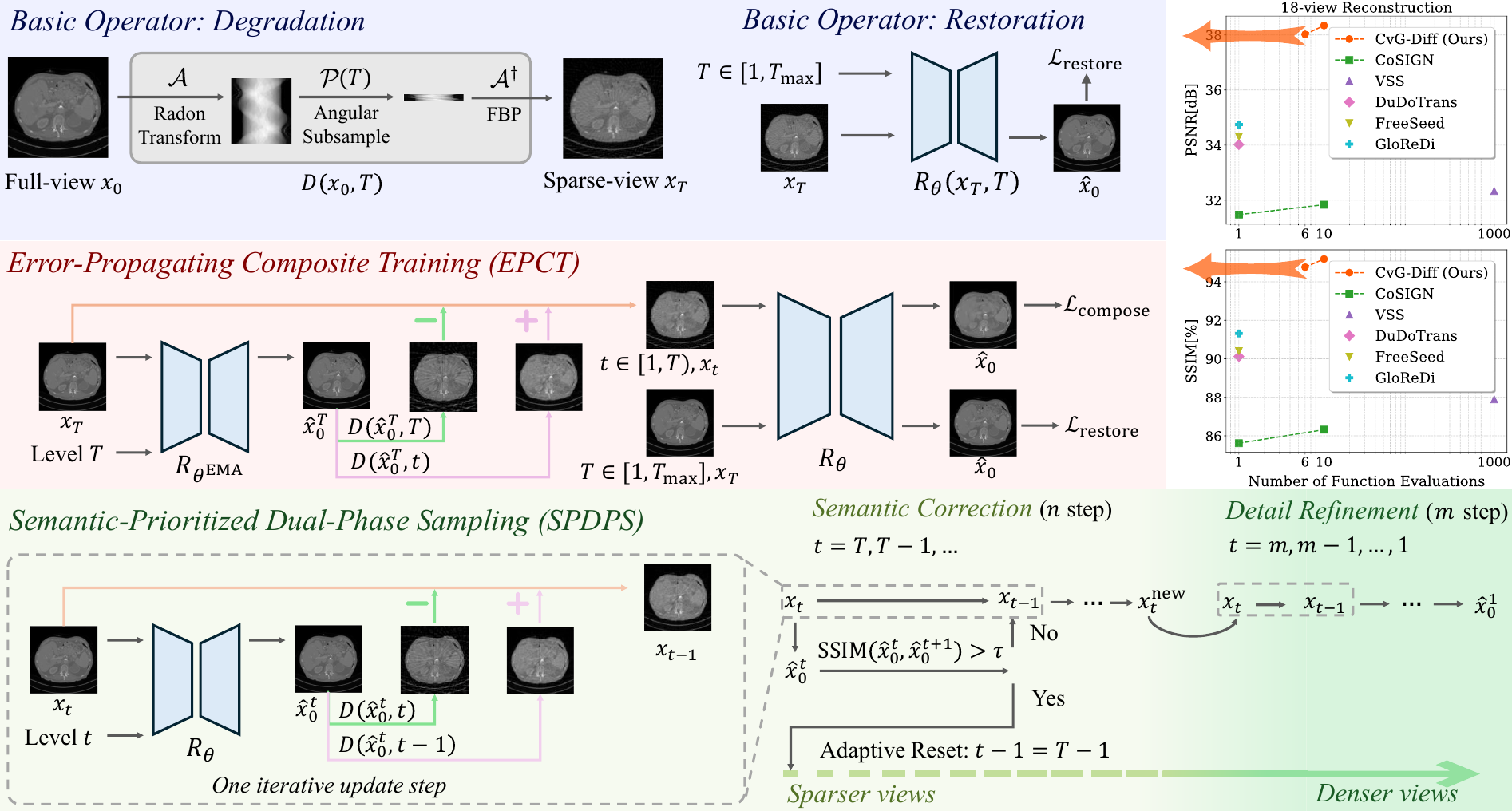}
    \caption{Overview of the proposed CvG-Diff.}
    \label{fig:overview}
\end{figure}

%
%
Despite these advancements, generalized diffusion models face two critical limitations in sparse-view CT reconstruction. First, intermediate reconstruction errors introduce propagated streak artifacts that models struggle to correct during iterative sampling. Second, conventional sequential sampling often dedicates excessive steps in refining anatomically stable results while under-correcting critical errors. To address these concerns, we propose an Error-Propagating Composite Training (EPCT) strategy, which simulates multi-step artifact accumulation during training to tackle propagated errors. Furthermore, we develop Semantic-Prioritized Dual-Phase Sampling (SPDPS), a strategy that prioritizes anatomical correctness before detail refinement. It adaptively resets to input sparse-view level to leverage improved intermediate reconstructions in identifying error-prone regions. As shown in Fig.~\ref{fig:overview}, these innovations enable CvG-Diff to achieve superior results with few sampling steps.

Our contributions are threefold: 1) We propose the CvG-Diff framework that explicitly models angular sparsity artifacts as a deterministic degradation process to harness the simultaneous reconstruction capability across different sparsity regimes, enabling high-quality iterative reconstruction with few sampling steps; 2) We develop the EPCT strategy that suppresses artifact propagation and the SPDPS that further improves reconstruction quality by prioritizing sparse-view level anatomical correctness over detail refinement; 3) Quantitative and qualitative results highlight CvG-Diff's advantages over various state-of-the-art sparse-view CT reconstruction methods.

%% file: docs/method.tex
\section{Methodology}
\subsection{Preliminary: Generalized Diffusion Model}
The generalized diffusion process~\cite{colddiff} establishes a framework for iterative image restoration through two core components: a degradation operator $D$ that corrupts an input image to various degradation levels, and a restoration operator $R$ trained to invert this degradation. Unlike traditional diffusion models that rely on stochastic Gaussian noise, $D$ can represent arbitrary degradation processes, provided it satisfies the boundary condition $D(x_0,0)=x_0$, where $x_0 \in \mathbb{R}^N$ denotes the clean image.
The forward process generates a degraded image $x_t$ at the severity level $t$ via $x_t = D(x_0,t)$. The restoration operator $R_\theta$, parameterized by a neural network with weights $\theta$, is then optimized to invert this degradation, achieving $R(x_t, t)\approx x_0$ by minimizing the following loss
\begin{equation}
    \mathcal{L}_{\text{restore}} = \| R_\theta(D(x_0, t), t) - x_0 \|_2.
    \label{loss1}
\end{equation}
During inference, the reverse sampling process iteratively refines degraded input $x_T$ at known degradation level $T$ by alternating between restoration and degradation operations.
Normally, the sampling schedule is the sequential reverse of training levels $t\in \{T,T-1,\ldots,1\}$, and one updating step is formulated as
\begin{align}
    x_{t-1} &= x_t - D(\hat{x}_0^t, t) + D(\hat{x}_0^t, t-1), \label{eq:degrade} \\
    \text{and } \hat{x}_0^t &= R_\theta(x_t, t). \label{eq:restore}
\end{align}
The final output is $\hat{x}_0 = \hat{x}_0^1 = R_\theta(x_1, 1)$. Formally, we denote the operation of iterative update using Eq.~(\ref{eq:degrade}) and Eq.~(\ref{eq:restore}) following a sequential schedule of $t \in \{T,T-1,\ldots,1\}$  as $I(x_T, T)$.

\begin{figure}[tbp]
    \centering
    \includegraphics[width=1.0\linewidth]{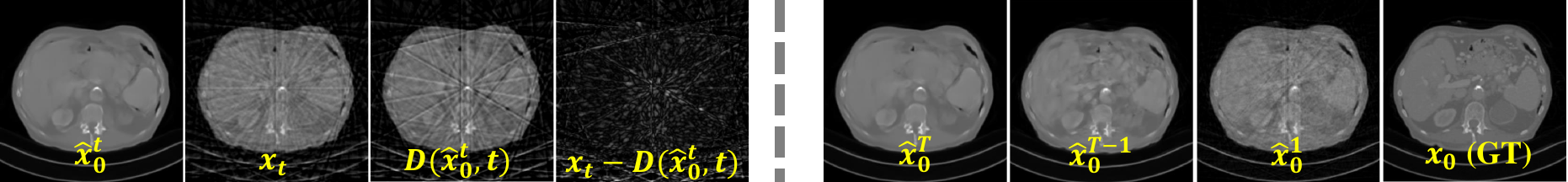}
    \caption{Error propagation issue in direct extension. \textbf{Left:} Incorrect reconstruction introduces streak artifacts. \textbf{Right:} Multi-step reconstruction results of performing $I(x_T,T)$. The model struggles to remove accumulated artifacts.}
    \label{fig:error_propagation}
\end{figure}
\subsection{Limitation of Extension to Sparse-view CT Reconstruction}
Sparse-view CT reconstruction aims to recover a high-quality image $x_0$ from subsampled projections of a degraded image $x_T$. To adapt the generalized diffusion framework to this task, CvG-Diff defines a deterministic degradation operator $D$ that explicitly models artifact patterns induced by angular subsampling: 
\begin{equation}
    x_T = D(x_0, T) = \mathcal{A}^\dagger\mathcal{P}(T)\mathcal{A}x_0,
    \label{eq:degrade_operation}
\end{equation}
where $\mathcal{A}$ denotes the Radon Transform, $\mathcal{P}(T)$ applies a subsampling mask to the sinogram at the severity level $T$, and $\mathcal{A}^\dagger$ represents FBP reconstruction back into image domain.
This formulation enables simultaneous training across multiple sparsity levels.
In practice, we define a severity level mapping $g(t)$ that assigns each discrete severity level $t$ to a number of views $\mathcal{T}_t$ from a sequence $\mathcal{T}$, spanning from the most-sparse-level $T_{\text{max}}$ to denser ones.
%



%
%
However, direct extension to sparse-view CT suffers from the inherent limitation of artifact propagation during iterative sampling. As shown in Fig~\ref{fig:error_propagation}, in each step, reconstruction errors within $\hat{x}_0^t$ contribute to additional streak artifacts when calculating $x_t - D(\hat{x}^t_0,t)$. Consequently, networks trained solely with Eq.~(\ref{loss1}) struggle to correct artifacts that deviate significantly from the current severity level, and the accumulated artifacts lead to subpar reconstructions.

\subsection{Cross-view Generalized Diffusion Model}
To address error propagation in multi-step sampling while maintaining computational efficiency, CvG-Diff integrates two key innovations: 1) an Error-Propagating Composite Training strategy (EPCT) that mitigates artifact accumulation, and 2) a Semantic-Prioritized Dual-Phase Sampling (SPDPS) mechanism that prioritizes anatomical correctness at sparse-view levels before detail enhancement.

\noindent \textbf{Error-Propagating Composite Training }
In each training step, we first update $R_\theta$ using Eq.~(\ref{loss1}) with a random target level $T \in [1, T_{\text{max}}]$. Then, we randomly select an intermediate level $t \in [1,T)$ to simulate the propagated artifacts from level $T$ to $t$. To this end, we maintain an exponential moving averaged (EMA) version of the restoration network $R_{\theta^{\text{EMA}}}$, where $\theta^{\text{EMA}}=\gamma\theta^{\text{EMA}} + (1-\gamma)\theta$ and is updated every $p$ training iterations. We use $R_{\theta^{\text{EMA}}}$ to stably generate reconstruction results at level $T$, and applies an additional loss as follows
\begin{align}
    \hat{x}_0^T &= R_{\theta^{\text{EMA}}}(x_T, T), \\
    x_{t} &= x_T -D(\hat{x}_0^T,T) + D(\hat{x}_0^T, t), \\
    \mathcal{L}_{\text{compose}} &= \|x_0 - R_\theta(x_{t}, t)\|_2.
    \label{loss2}
\end{align}
We use Eq.~(\ref{loss2}) to further update $R_\theta$, which is enforced to address potential artifacts introduced in level $t$ and identifies error-prone regions for correction. By this way, CvG-Diff learns to handle artifacts propagated from different levels, enabling reliable reconstruction with larger inter-step subsampling gaps.

\noindent \textbf{Semantic-Prioritized Dual-Phase Sampling}
During multi-step inference, reconstruction steps at sparse-view levels often introduce over-smoothed errors that blur anatomical boundaries. While reconstruction steps at subsequent denser-view levels refine high-frequency structures, they struggle to fully rectify these significant blurred regions. Therefore, an optimal restoration strategy should prioritize anatomical semantics at sparse-view levels before enhancing finer structures.
However, the fixed progression steps in conventional sequential sampling fail to adequately satisfy such requirements (see example in Fig.~\ref{fig:spdps_example}).

To this end, given $N$ sampling steps, we partition $N=n+m$ into first $n$ steps for a semantic correction phase, followed by $m$ steps for a detail refinement phase. For semantic correction, we start with performing $I(x_T,T)$, incorporating an adaptive resetting criterion based on structural similarity (SSIM) comparisons. When updating at $t$-step, we evaluate whether $\text{SSIM}(\hat{x}_0^t, \hat{x}_0^{t+1}) > \tau$. If the condition is satisfied, we reset to input sparse-view level by
\begin{equation}
    x_{T-1}' = x_T - D(\hat{x}_0^{t}, T) + D(\hat{x}_0^t, T-1). \label{eq:sqdps_eq}
\end{equation}
The rationale behind Eq.~(\ref{eq:sqdps_eq}) is that once anatomical convergence is detected, we obtain an improved reconstruction result $\hat{x}_0^{t}$ over $\hat{x}_0^{T}$ without wasting excessive steps, and the comparison between $D(\hat{x}_0^{t}, T)$ and $x_T$ helps to better identify error-prone regions, improving anatomical accuracy at sparser-view levels. Then, we use $x_{T-1}'$ to perform $I(x_{T-1}',T-1)$. This adaptive mechanism repeats until updating $n$ steps, and we obtain $x_{t}^{\text{new}}$. Afterwards, for detail refinement, we conduct $\hat{x}_0=I(x_{t}^{\text{new}}, m)$ at denser-view levels to obtain the final result.

%% file: docs/experiments.tex
\section{Experiments}
\begin{table}[tbp] 
    \centering
    \caption{Comparison results on AAPM-LDCT. Mean results of RMSE [HU], PSNR [dB], SSIM [$\%$], and reconstruction time [s] are reported. NFE denotes the number of network function evaluation ($N$) in multi-step reconstruction. The best result is shown in \textbf{bold} and the second-best results are \underline{underlined}.}
    \label{tab:sota}
    \resizebox{\textwidth}{!}{
    \begin{tabular}{l|c|c|c|c|c|c|c|c|c|c}
    \hline
        \multirow{2}{*}{Method} & \multicolumn{3}{c|}{18-view} & \multicolumn{3}{c|}{36-view} & \multicolumn{3}{c|}{72-view} & \multirow{2}{*}{Time} \\
    \cline{2-10}
     & RMSE$\downarrow$ & PSNR$\uparrow$ & SSIM$\uparrow$ & RMSE$\downarrow$ & PSNR$\uparrow$ & SSIM$\uparrow$ & RMSE$\downarrow$ & PSNR$\uparrow$ & SSIM$\uparrow$ & \\
    \hline
    \multicolumn{11}{c}{One-step feed-forward methods}  \\
    \hline
    DudoTrans~\cite{dudotrans} & 58.83 & 34.02 & 90.12 & 37.78 & 38.24 & 94.13 & 20.56 & 42.76 & 97.62 & 0.13 \\
    Freeseed~\cite{freeseed} & 58.09 & 34.31 & 90.40 & 38.26 & 37.92 & 93.93 & 21.47 & 42.93 & 97.53 & \underline{0.07} \\
    Glorei~\cite{glorei} & 57.03 & 34.75 & 91.32 & 39.98 &  37.54 & 93.52 & 27.85 & 41.80 & 96.30 & \textbf{0.06} \\
    \hline
     \multicolumn{11}{c}{Multi-step diffusion-based methods} \\
    \hline
    VSS~\cite{vss} (NFE=1000) & 72.62 & 32.34 & 87.90 & 40.28 & 37.52 & 93.99 & 24.14 & 41.92 & 97.07 & 264.71 \\
    CoSIGN~\cite{cosign} (NFE=1) & 80.24 & 31.48 & 85.62 & 62.21 & 33.68 & 88.71 & 49.82 & 35.63 & 91.40 & 0.09 \\
    CoSIGN~\cite{cosign} (NFE=10) & 76.95 & 31.84 & 86.31 & 53.73 & 34.96 & 89.67 & 38.37 & 37.87 & 93.20 & 1.66 \\
    CvG-Diff (Ours) (NFE=6) & \underline{37.97} & \underline{38.02} & \underline{94.76} & \underline{24.98} & \underline{41.63} & \underline{96.92} & \underline{15.70} & \underline{45.67} & \underline{98.54} & 0.39 \\
    CvG-Diff (Ours) (NFE=10) & \textbf{36.63} & \textbf{38.34} & \textbf{95.18} & \textbf{24.63} & \textbf{41.78} & \textbf{97.05} & \textbf{15.18} & \textbf{45.94} & \textbf{98.63} & 0.68 \\
    \hline
    \end{tabular}
    }
\end{table}
\subsection{Experiment Settings}
We evaluate our method on the AAPM Low-Dose CT (LDCT) dataset~\cite{aapm16}, which contains 5,936 CT slices from 10 patients. We split the data into 5,410 training/validation slices (9 patients, 9:1 split) and 526 test slices (1 patient). Sparse-view CT projections are simulated using a fan-beam geometry with a source-to-detector distance of 59.5 cm, 672 detector elements, and scan parameters of 120 kVp and 500 mA, implemented via the TorRadon toolbox~\cite{ronchetti2020torchradon}. To assess robustness across sparsity levels, we reconstruct images from $N_v \in \{18, 36, 72\}$ projection views.

Our reconstruction network adopts a Diffusion UNet architecture with residual blocks, a base feature dimension of 128, and channel multipliers $[1,2,2,2]$ across four resolution scales~\cite{colddiff}. The model is trained for 40 epochs with a batch size of 4 using the Adam optimizer $(\beta_1 = 0.9,\beta_2 = 0.999)$ with an initial learning rate of $4\times 10^{-5}$, reduced by a factor of 0.8 after epoch 25. The EMA model is updated by setting $\gamma=0.995$ and $p=10$.
We set $\mathcal{T}=[288, 234, 180, 126, 72, 54, 36, 18]$ that covers all three target numbers of views (\ie, 18, 36, 72), such that only one model needs to be trained to handle all our experimental settings.
For SPDPS strategy, we set $\tau=0.97$ and $m=4$. Experiments are conducted on a single NVIDIA 3090 GPU. All sparse-view CT reconstruction methods are evaluated quantitatively in terms of root mean squared error (RMSE) in Hounsfield Units (HU), peak signal-to-noise ratio (PSNR), and structural similarity (SSIM). 

\begin{figure}[tbp]
    \centering
    \includegraphics[width=1.0\linewidth]{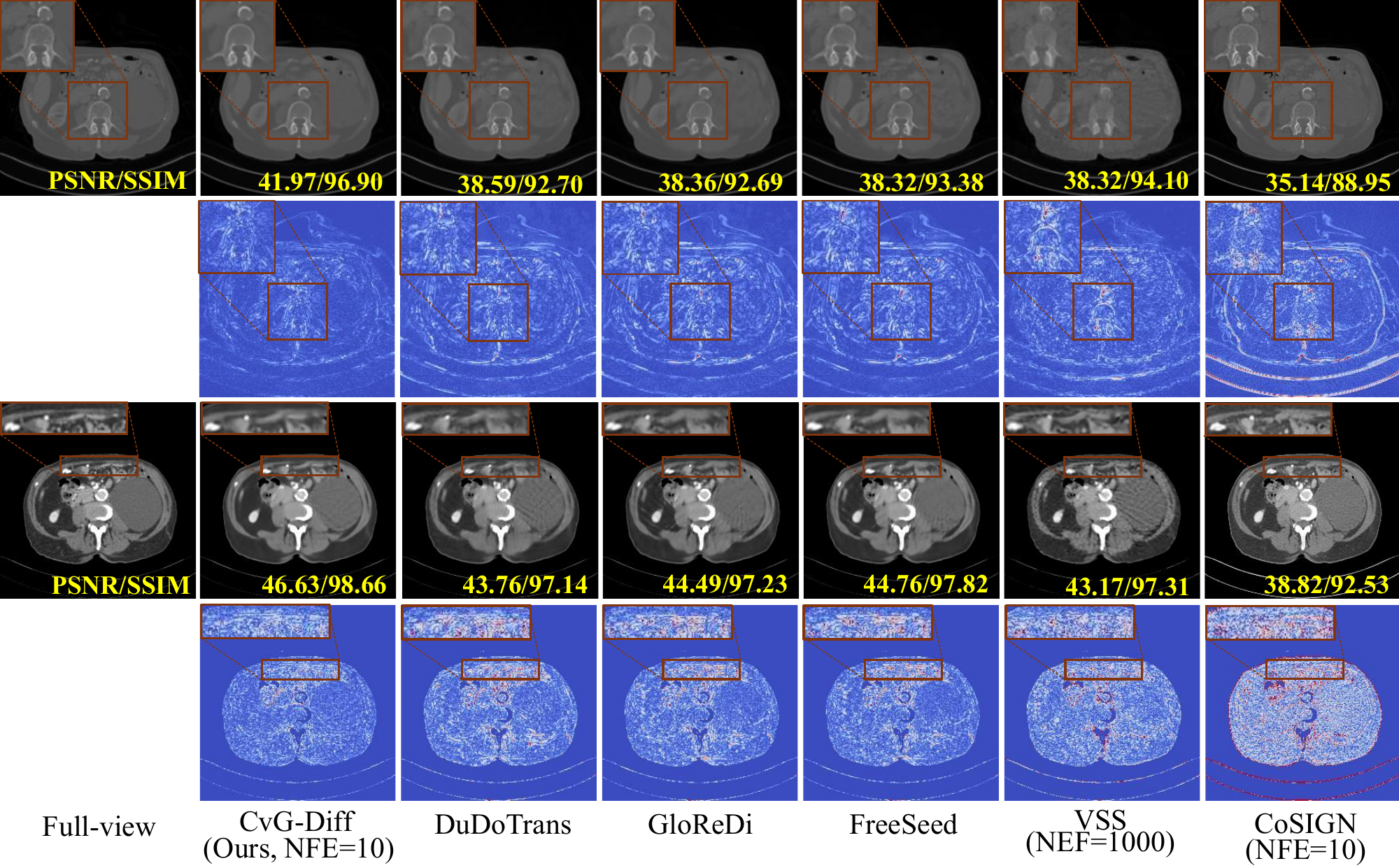}
    \caption{Visual comparison of different methods. From top to bottom: $N_v=\{36,72\}$ with display windows $[-1000, 2000], [-200,300]$ HU, respectively. Red color in error maps indicate a larger error.}
    \label{fig:qualitative}
\end{figure}

\subsection{Comparison with State-of-The-Arts}

We evaluate CvG-Diff against two categories of state-of-the-art sparse-view CT reconstruction methods: (1) one-step feed-forward methods (FreeSeed~\cite{freeseed}, GloReDi~\cite{glorei}, DuDoTrans~\cite{dudotrans}), which train networks on paired sparse-view and full-view data for artifact suppression at specific subsampling rates; (2) multi-step diffusion-based methods (VSS~\cite{vss}, CoSIGN~\cite{cosign}), which iteratively refine reconstructions using generative priors on full-view data.

Quantitative results in Tab.~\ref{tab:sota} demonstrate that one-step methods outperform diffusion baselines due to artifact-specific optimization. On the other hand, diffusion frameworks like VSS achieve comparable results at higher sampling rates (e.g., 72-view) using a unified model that bypasses paired data training, albeit at the cost of prohibitively long sampling steps. CoSIGN mitigates this inefficiency through consistency distillation on paired data, reducing sampling steps while retaining multi-step refinement capabilities. Nevertheless, both diffusion-based approaches struggle under extreme sparsity (e.g., 18-view), where insufficient projection measurements amplify reconstruction errors. In contrast, CvG-Diff consistently outperforms all baselines by leveraging cross-view reconstruction capability across subsampling levels, achieving superior results at all tested sparsity regimes (18/36/72-view) with $\le 10$ sampling steps and comparable inference time. Qualitative results in Fig.~\ref{fig:qualitative} highlight the advantages of CvG-Diff. In high-sparsity cases, one-step methods produce over-smoothed outputs, while diffusion-based methods reconstruct visually vivid structures with low fidelity. Instead, CvG-Diff successfully recovers precise anatomical structures. 

\subsection{Ablation Study}
\begin{figure}[t]
    \centering
    \includegraphics[width=\linewidth]{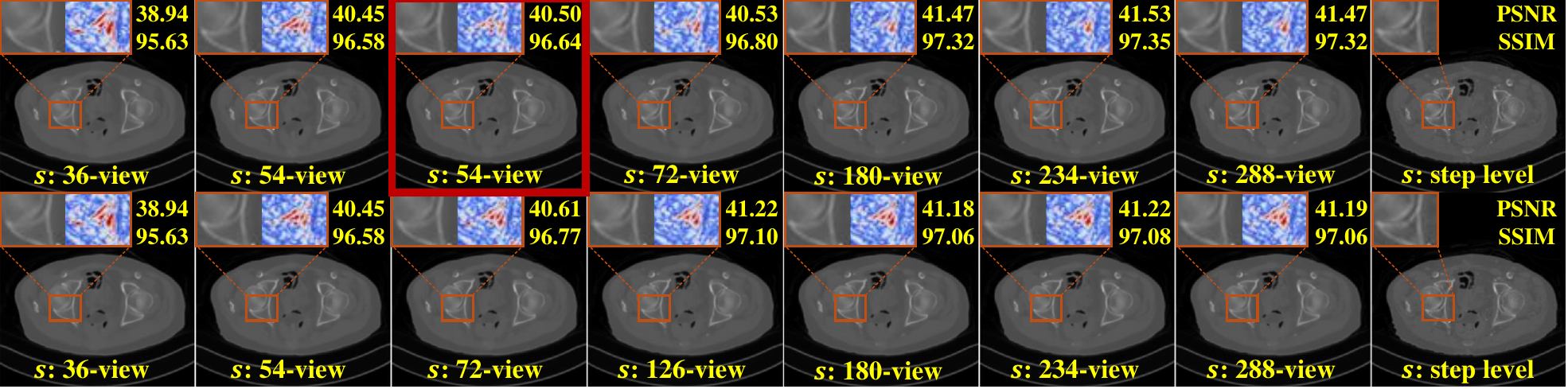}
    \caption{Visualization of multi-step reconstruction results between SPDPS (top) and sequential sampling (bottom). Error maps show deviations from ground truth (last column) in zoomed region, and red color indicates larger errors. The red box denotes the step when SPDPS resets the degradation level, which rectifies blurred boundaries.}
    \label{fig:spdps_example}
\end{figure}

\begin{table*}
        [!t]
        \begin{minipage}[t]{0.48\textwidth}
            \makeatletter\def\@captype{table}\makeatother
            \centering
            \caption{Ablation for EPCT and SPDPS.}
            \label{tab:ablation}
        
            \resizebox{\textwidth}{!}{
            \begin{tabular}{ccc|c|c|c}
    \hline
    \multicolumn{3}{c|}{Settings} & \multicolumn{1}{c|}{18-view} & \multicolumn{1}{c|}{36-view} & \multicolumn{1}{|c}{72-view}  \\
    ID & EPCT & SPDPS & PSNR$\uparrow$/SSIM$\uparrow$ & PSNR$\uparrow$/SSIM$\uparrow$ & PSNR$\uparrow$/SSIM$\uparrow$ \\
    \hline
    A) & \ding{55} & \ding{55} & 33.67/82.42 & 37.02/89.15 & 43.38/97.30 \\ 
    B) & \ding{55} & \ding{51} & 34.67/85.23 & 38.23/92.07 & 43.23/97.54 \\
    C) & \ding{51} & \ding{55} & 37.85/94.68 & 41.48/96.87 & 45.66/98.54 \\
    Ours & \ding{51} & \ding{51} & \textbf{38.34}/\textbf{95.18} & \textbf{41.78}/\textbf{97.05} &  \textbf{45.94}/\textbf{98.63}\\
    \hline
    \end{tabular}
            }
        \end{minipage}\quad
        \begin{minipage}[t]{0.48\textwidth}
            \makeatletter\def\@captype{table}\makeatother
            \centering
            \caption{Influence of $\tau$ and $m$ for SPDPS.}
            \label{tab:sensitivity} \resizebox{\textwidth}{!}{
            \begin{tabular}{c|c|c|c|c|c|c|c}
    \hline
    \multicolumn{1}{c|}{Parameter} & \multicolumn{3}{c|}{18-view} & \multicolumn{1}{|c|}{Parameter} & \multicolumn{3}{c}{18-view} \\
    \hline
    $\tau$ & \multicolumn{3}{c|}{PSNR$\uparrow$/SSIM$\uparrow$} & $m$ & \multicolumn{3}{c}{PSNR$\uparrow$/SSIM$\uparrow$} \\
    \hline
    0.97 & \multicolumn{3}{c|}{\textbf{38.34}/\textbf{95.18}} & 2 & \multicolumn{3}{|c}{38.33/95.13} \\ 
    0.98 & \multicolumn{3}{c|}{38.06/95.03} & 3 & \multicolumn{3}{|c}{\textbf{38.34}/95.15} \\
    0.99 & \multicolumn{3}{c|}{37.94/94.96} & 4 & \multicolumn{3}{|c}{\textbf{38.34}/\textbf{95.18}} \\
    \hline
    \end{tabular}
            }
        \end{minipage}\quad
    \end{table*}

We validate our proposed strategies by comparing three CvG-Diff variants: A) The baseline implementation of the generalized diffusion framework using degradation operator defined in Eq.~(\ref{eq:degrade_operation}); B) Variant A + SPDPS inference strategy; C) Variant A + EPCT strategy during training and uses sequential sampling during inference. 

Quantitative results in Tab.~\ref{tab:ablation} demonstrate the incremental benefits of our innovations. While variant B improves over variant A, the limited performance indicates that enhancing inference efficiency alone cannot address the core challenge of artifact propagation. Variant C resolves this limitation through EPCT, which simulates multi-step degradation during training to suppress error accumulation, yielding a significant 3.80 dB averaged PSNR improvement over variant A. The best performance is achieved by combining both EPCT with SPDPS.
Fig.~\ref{fig:spdps_example} compares variant C and D, demonstrating that SPDPS corrects blurred anatomical boundaries by adaptively resetting degradation levels, enhancing subsequent detail refinement. In contrast, sequential sampling propagates uncorrected errors. Further analysis in Tab.~\ref{tab:sensitivity} validates the robustness of SPDPS across different $\tau$ and $m$ configurations, with $\tau$ (degradation reset) impacting performance more significantly than $m$ (refinement steps). 

%% file: docs/conclusion.tex
\section{Conclusion}

In this study, we present CvG-Diff, a novel framework for sparse-view CT reconstruction based on the generalized diffusion process. By addressing two critical limitations regarding the artifact propagation across iterative sampling and inefficient correction of over-smoothed artifacts at sparse-view levels, CvG-Diff achieves superior reconstruction results with few ($\le$ 10) sampling steps across various sparse-view scenarios. Our findings demonstrate the benefits of harnessing the simultaneous reconstruction capability across different sparse-view levels. Future research could explore dual-domain generalized diffusion models that jointly optimize sinogram and image-domain reconstructions and the corresponding optimal sampling strategy with collaborative efforts in both domains. Further investigation of this direction will be a focus of our future work.

\subsubsection{\ackname}
This work was partially supported by grants from the Hong Kong Innovation and Technology Fund under Project PRP/041/22FX.

\subsubsection{Disclosure of Interests.}
The authors have no competing interests to declare that are relevant to the content of this article.